\documentclass[a4paper,fleqn]{cas-sc}

\usepackage[utf8]{inputenc}   % for older LaTeX

\usepackage{amsmath,amssymb,amsfonts}%
\usepackage{amsthm}%
\usepackage{mathrsfs}%
\usepackage[title]{appendix}%
\usepackage{xcolor}%
\usepackage{textcomp}%
\usepackage{manyfoot}%
\usepackage{booktabs}%
\usepackage{algorithm}%
\usepackage{algorithmicx}%
\usepackage{algpseudocode}%
\usepackage{listings}%

\usepackage{capt-of}

\usepackage{url}
\usepackage{graphicx}% Include figure files
\usepackage{dcolumn}% Align table columns on decimal point
\usepackage{bm}% bold math
\usepackage{tabularx}
\usepackage{amssymb}
\usepackage{amsmath}
\usepackage{stackengine}
\usepackage{pgfplots}
\usepackage{makecell, multirow, tabularx}
\usepackage{array}
\usepackage{booktabs}
\usepackage{mathtools}

\usepackage[section]{placeins}
\usepackage{enumitem}
\usepackage{subcaption}
\pgfplotsset{compat=1.18}
\usepackage[utf8]{inputenc}
\usepackage[T1]{fontenc}
\usepackage{etoolbox}
\usepackage{siunitx} 
\usepackage{adjustbox} %\usepackage{multirow}   % multirow cells
\usepackage{xcolor}

\usepackage[numbers,sort&compress]{natbib}
\def\tsc#1{\csdef{#1}{\textsc{\lowercase{#1}}\xspace}}
\tsc{WGM}
\tsc{QE}
\tsc{EP}
\tsc{PMS}
\tsc{BEC}
\tsc{DE}
\begin{document}
\let\WriteBookmarks\relax
\def\floatpagepagefraction{1}
\def\textpagefraction{.001}

% Short title
\shorttitle{Scaling laws in Stablecoin}

% Short author
%\shortauthors{Kundan Mukhia et~al.}

% Main title of the paper
\title [mode = title]{Scaling laws of Stablecoin Transactions: Evidence from USDT and USDC on the Ethereum blockchain}                      
% Title footnote mark
% eg: \tnotemark[1]
%\tnotemark[1,2]

% Title footnote 1.
% eg: \tnotetext[1]{Title footnote text}
% \tnotetext[<tnote number>]{<tnote text>} 
%\tnotetext[1]{This document is the results of the research project funded by the National Science Foundation.}

%\tnotetext[2]{The second title footnote which is a longer text matter  to fill through the whole text width and overflow into another line in the footnotes area of the first page.}

% First author
%
% Options: Use if required
% eg: \author[1,3]{Author Name}[type=editor,
%       style=chinese,
%       auid=000,
%       bioid=1,
%       prefix=Sir,
%       orcid=0000-0000-0000-0000,
%       facebook=<facebook id>,
%       twitter=<twitter id>,
%       linkedin=<linkedin id>,
%       gplus=<gplus id>]
% =========================================================
% ADDITIONAL AUTHORS (DO NOT MODIFY EXISTING FORMAT ABOVE)
% =========================================================

% Fifth author
\author[1]{Kundan Mukhia}
\ead{kundanmukhia07@gmail.com}
\credit{Writing – original draft, Visualization, Validation, Software, Methodology, Investigation, Formal analysis, Conceptualization}

% Sixth author
\author[1]{Sabat Rai}
\ead{sabatrai2105@gmail.com}
\credit{Writing – review \& editing, Investigation}

% Seventh author
\author[2]{Vivek Shrivastav}
\ead{vivekshrivastav1998@gmail.com}
\credit{Writing –review \& editing, Investigation}

\author[3]{Imran Ansari}
\ead{imranansari@iisc.ac.in}
\credit{Writing – review \& editing,  Validation}

% Eighth author
\author[1]{Md. Nurujjaman \corref{cor1}}
\ead{md.nurujjaman@nitsikkim.ac.in}
\credit{Writing – review \& editing, Visualization, Validation, Supervision, Formal analysis, Conceptualization}

% Ninth author

% ---------------------------------------------------------
% AFFILIATIONS
% ---------------------------------------------------------

\affiliation[1]{organization={National Institute of Technology Sikkim},
    addressline={Department of Physics},
    city={Ravangla},
    state={Sikkim},
    postcode={737139},
    country={India}}

\affiliation[2]{
    organization={Department of Physics, Sikkim University},
    city={Gangtok},
    postcode={737102},
    state={Sikkim},
    country={India}
}

\affiliation[3]{organization={Department of Management Studies, Indian Institute of Science},
   % addressline={CV Raman Road},
    city={Bengaluru},
   % state={Karnataka},
    postcode={560012},
    country={India}}

\cortext[cor1]{Corresponding author}
% ---------------------------------------------------------
% Corresponding author text
% ---------------------------------------------------------
%\cortext[cor3]{Corresponding author}

% Footnote text

\begin{abstract}
Stablecoins have rapidly emerged as a major class of important digital assets and an important component of the digital financial ecosystem. Despite their growing importance, the statistical properties of stablecoin transaction activity remain largely unexplored. To the best of our knowledge, this is the first study to investigate the scaling behavior in stablecoin transaction blockchain data, focusing on USDT and USDC. We analyzed approximately 370 million USDT and USDC transaction data recorded on the Ethereum blockchain across six periods spanning June 2024 to February 202. Based on the interaction categories between Externally Owned Accounts (EOAs) and Smart Contracts (SCs), we classified the transactions into four categories: EOA–EOA, EOA–SC, SC–EOA, and SC–SC. Using maximum-likelihood estimation of power-law exponents, we find that transaction value distributions exhibit heavy-tailed scaling across both stablecoins in all periods and in all four interaction categories. Across the four interaction categories, the estimated exponent fall within $\alpha \approx 1.4$-$1.8$, consistent with values previously reported for some heavy-tailed financial quantities. Within this overall range, we identify two distinct scaling regimes: the EOA-involved categories cluster exponents around $\alpha \approx 1.45$-$1.60$, whereas SC-SC transactions exhibit higher exponents of approximately $\alpha \approx 1.72$-$1.73$. A sensitivity analysis further confirms that this separation is robust across all six periods, stablecoins, and fitting sample sizes. Additionally, our counterfactual analysis shows that changes in category weights alone cannot explain the magnitude of the observed variations in the overall exponent. Using different sample sizes, we show that the counterfactual path accounts for only about $10\%$-$35\%$ of the total temporal range observed in the actual data. Overall, our results indicate two broadly differentiated, interaction-associated scaling regimes in the tail of distributions of stablecoin transaction value. The power-law tail behaviour is observed throughout the stablecoin transaction activity, but the exponent depends on whether the transaction is driven by EOAs or SCs. Our findings provide a basis for further research on scaling behaviour and transaction heterogeneity into blockchain-based financial systems.

\end{abstract}


\maketitle

\section{Introduction}
\label{introduction}

In recent years, stablecoins have evolved from a specialized cryptocurrency instrument into an important component of the digital financial ecosystem. Designed to maintain a stable value relative to a reference asset, most commonly the U.S. dollar, they are widely used for trading, settlement, decentralized finance (DeFi), liquidity provision, and cross-border value transfer \cite{arner2020stablecoins,catalini2022some, mukhia2025early, lyons2023keeps, duan2023instability}. In 2025, the global stablecoin market reached a total capitalization of more than US~\$300 billion, with a transaction volume of approximately US~ \$33-35 trillion \cite{Carapella2026Stablecoins, Bloomberg2026Stablecoin}. The transaction activity of the stablecoins within the Ethereum blockchain involves two categories of accounts, namely the Externally Owned Accounts
(EOAs) and Smart Contracts (SCs)~\cite{mukhia2025universal, chen2020traveling, kolvart2016smart, szabo1997idea}. The interactions between these account types create heterogeneous transaction dynamics. The Ethereum blockchain records these transfers with detailed temporal and monetary information, which provides an opportunity to study the statistical organization of blockchain-based financial activity.

In statistical physics and econophysics, scaling laws and universality are among the most fundamental concepts that are successfully applied to understand complex systems. In financial markets, the collective actions of heterogeneous participants generate heavy-tailed distributions and other statistical patterns that remain stable across assets, markets, and time periods \cite{gopikrishnan1998inverse, stanley2008statistical, sato2025strict, mukhia2026core, mukhia2024complex, plerou1999scaling}.  Determining whether such patterns to follow scaling laws or are dependent on market structure and interaction mechanisms is therefore an important question in the statistical physics of financial systems. This leads to the question of whether stablecoin transaction values follow the same statistical scaling patterns observed in traditional financial markets or whether their scaling behaviour depends on the interaction mechanism.

Extensive empirical evidence for scaling has been reported in both traditional financial and cryptocurrency markets. In stock markets, quantities such as returns, trading volume, trade size, intertrade times, and price impact have all been shown to exhibit power-law behaviour. Plerou and Stanley \cite{plerou2008stock} found inverse-cubic tails in stock-return distributions across the U.S., London, and Paris markets, with exponents $\zeta_R \approx 3$. Ivanov et al. \cite{ivanov2004common} reported common scaling patterns in intertrade times across companies from different economic sectors. Trade sizes and share volumes have also been shown to follow power-law distributions with exponents around $\zeta_q \approx 1.5$, and $\zeta_Q \approx 1.7$ \cite{stanley2008statistical}. Zhou et al. \cite{zhou2005inverse} showed that the probability distribution of exit times follows a universal power law with exponent $\alpha \approx 1.5$ across both developed and emerging markets. More recently, Sato and Kanazawa \cite{sato2025strict} presented high-precision evidence for the universality of the square-root price-impact law across stocks. Several studies have also shown a similar scaling pattern in the cryptocurrency markets. Begušić et al. \cite{beguvsic2018scaling} reported power-law scaling in Bitcoin return distributions across multiple exchanges with exponents in the range $2 < \alpha < 2.5$, indicating heavier tails than those observed in stock returns.  Wu et al. \cite{wu2018classification} also identified power-law behaviour in cryptocurrency market capitalizations. Similarly, Wątorek et al. \cite{wkatorek2021multiscale} showed that cryptocurrency markets exhibit many statistical properties associated with mature financial systems, including heavy-tailed fluctuations, long-range correlations, fractality, and multiscaling. Recent studies on ERC20 token transfers on Ethereum have also shown to follow power-law scaling behaviour\cite{mukhia2025universal}. At the transaction level, Li et al. \cite{li2019exponentially} found multiple scaling regimes in Bitcoin trade-size distributions. Therefore, these studies indicate that heavy-tailed behaviour is widespread in both traditional and cryptocurrency markets, but the stability of the exponent can depend on the asset, market, period, and quantity being analysed.

Most research on stablecoins has primarily focused on price stability, reserve mechanisms, adoption, and financial stability \cite{arner2020stablecoins,catalini2022some,duan2023instability, mahrous2026stablecoins,berentsen2019stablecoins,klages2020stablecoins}. Despite the rapid growth of stablecoin usage in digital finance, the scaling properties of their transfer values remain largely unknown. To fill this gap, we examine scaling behaviour in stablecoin transfers. We focus on Tether (USDT) and USD Coin (USDC), which together account for approximately $85\%$ of stablecoin market capitalization \cite{claessens2026papers,mukhia2025early, coinmarketcap2025stablecoin}. Stablecoin transactions on Ethereum involve interactions between EOAs and SCs. This provides a systematic basis for testing whether the statistical properties of transaction values depend on the underlying interaction mechanism.

In this study, we examine approximately 370 million USDT and USDC transfers on the Ethereum blockchain over six periods to analyse scaling behaviour in stablecoin transaction values. We classify the transactions into four interaction categories based on the account types involved: EOA--EOA, EOA--SC, SC--EOA, and SC--SC. We also consider the overall transaction data, which combines all four interaction categories without distinguishing between the account types. For each category and the overall distribution, we estimate power-law exponents and assess their robustness across all periods and sample sizes. We then perform a counterfactual analysis to test whether shifts in category weights can account for the temporal variation in the Overall exponent. We find that a power-law tail behaviour form is present across all categories, as observed in traditional financial markets. But the exponent varies between the EOA-involved transactions and SC-SC transactions, and these differences hold across all periods and samples. Taken together, our results suggest a power-law tail behaviour throughout the stablecoin transaction values, but the exponent depends on whether the transaction is driven by EOAs or SCs.

The remainder of the paper is organized as follows. Section~\ref{sec: Method of analysis} presents the methodology, followed by a description of the dataset in Section~\ref{sec:data_description}. The empirical results are presented in Section~\ref{sec:Result}, and Section~\ref{sec:Conclusion} concludes the paper.

\section{Method of analysis}
\label{sec: Method of analysis}

To study the scaling behaviour of stablecoin transaction values on the Ethereum blockchain, we apply several methods. We first estimate power-law form exponents for the stablecoin transaction value using maximum-likelihood estimation. To assess robustness, we conduct a sensitivity analysis across five sample sizes and random seeds. Finally, we apply a composition counterfactual analysis to see whether changes in category weights can explain the temporal variation in the overall exponent. Further more details on each method are discussed below.

\subsection{Estimation of power law tail exponents}

Heavy-tailed distributions are a well-established characteristic of complex financial systems, in which extreme events occur substantially more frequently than predicted by conventional statistical distributions~\cite{clauset2009power, mitzenmacher2004brief,barabasi1999emergence,newman2005power}. Power-law scaling has been reported for a wide range of financial variables, including asset returns, trading volume, trade size, volatility, and foreign-exchange fluctuations~\cite{lux2016financial,bouchaud2001power}. For a continuous random variable, the probability density function for the scaling region is expressed as~\cite{clauset2007power}:
\(\displaystyle
p(x) = C x^{-\alpha},
\quad x \geq x_{\min},
\) where $C$ is a normalization constant, $x_{\min}$ denotes the lower bound of the scaling region, and $\alpha$ is the tail exponent. The corresponding complementary cumulative distribution function (CCDF) is

\begin{equation}
P(X \geq x)
=
\left(\frac{x}{x_{\min}}\right)^{-(\alpha-1)},
\qquad x \geq x_{\min}.
\label{eq:powerlaw_ccdf}
\end{equation}
Estimating the tail exponent requires identifying the region over which power-law scaling is observed. Following the general framework of Clauset et al.~\cite{clauset2007power}, the lower threshold, $x_{\min}$, is selected by minimizing the Kolmogorov--Smirnov (KS) distance between the empirical and fitted cumulative distributions. For computational efficiency, the threshold search is performed over a fixed set of candidate values covering the relevant range of the empirical distribution. The KS distance is defined as~\cite{goldstein2004problems,berger2014kolmogorov,massey1951kolmogorov}

$$
D =
\sup_{x \geq x_{\min}}
\left| S(x) - P_{\mathrm{PL}}(x) \right|,
$$where $S(x)$ is the empirical cumulative distribution function and $P_{\mathrm{PL}}(x)$ is the cumulative distribution function of the fitted power-law model. The value that produces the minimum KS distance is selected as $\hat{x}_{\min}$.

The estimation is performed in two stages. First, $\hat{x}_{\min}$ is determined from a randomly selected fitting dataset by minimizing the Kolmogorov--Smirnov distance between the empirical and fitted CCDFs. Second, conditional on the estimated threshold, the tail exponent is calculated using all available observations satisfying $x_i \geq \hat{x}_{\min}$. For a continuous power-law distribution, the maximum-likelihood estimator is~\cite{barndorff1996prediction,wasserman2004all}

$$
\hat{\alpha}
=
1+
n_{\mathrm{tail}}
\left[
\sum_{i=1}^{n_{\mathrm{tail}}}
\ln
\left(
\frac{x_i}{\hat{x}_{\min}}
\right)
\right]^{-1},
$$

where $n_{\mathrm{tail}}$ is the total number of observations in the complete dataset satisfying $x_i \geq \hat{x}_{\min}$. Conditional on $\hat{x}_{\min}$, the asymptotic standard error of the estimated exponent is calculated as $\displaystyle\sigma_{\hat{\alpha}}
= \frac{\hat{\alpha}-1}{\sqrt{n_{\mathrm{tail}}}},$ and the corresponding approximate 95\% confidence interval is $\mathrm{CI}_{95\%}=\hat{\alpha}\pm1.96\sigma_{\hat{\alpha}}.$

These confidence intervals quantify the uncertainty in $\hat{\alpha}$ conditional on the selected threshold and do not include the additional uncertainty associated with estimating $x_{\min}$. Compared with ordinary least-squares fitting on log--log coordinates, maximum-likelihood estimation generally provides more reliable and less biased estimates of power-law exponents.

\subsubsection{Sensitivity and temporal stability analysis}
\label{sec:Method_Sensitivity}

To evaluate the robustness of the estimated tail exponent, $\hat{\alpha}$, the estimation procedure is repeated for five fitting sample sizes, $n_{\mathrm{fit}}
\in
\left\{
50{,}000,\,
100{,}000,\,
200{,}000,\,
400{,}000,\,
800{,}000
\right\}$ and five independent random seeds, $s \in \{1,2,3,4,5\}$. This analysis allows us to examine whether the estimated exponent and lower scaling threshold, $\hat{x}_{\min}$, are sensitive to the fitting sample size or random selection of observations.

For each fitting size, the seed-averaged exponent is calculated as

\begin{equation}
\overline{\alpha}(n_{\mathrm{fit}})
=
\frac{1}{5}
\sum_{s=1}^{5}
\hat{\alpha}_{s}(n_{\mathrm{fit}})
\label{eq:mean_alpha_seeds}
\end{equation}

while the corresponding between-seed variability is measured by $\sigma_{\alpha}(n_{\mathrm{fit}})  =\sqrt{ \frac{1}{4} \sum_{s=1}^{5} \left[\hat{\alpha}_{s}(n_{\mathrm{fit}}) - \overline{\alpha}(n_{\mathrm{fit}})\right]^2
}
$

A small between-seed variation and convergence of $\overline{\alpha}$ as $n_{\mathrm{fit}}$ increases indicate that the estimated scaling exponent is robust to random variation and fitting sample size. Conversely, substantial variation would suggest sensitivity to the selected observations or fitting conditions.

Temporal stability is examined by dividing the transaction data into consecutive time periods and estimating $\hat{\alpha}$ separately for each stablecoin and interaction category. This enables us to determine whether the scaling behaviour remains stable over time and across different types of network interactions. Consistent exponent values across fitting sizes, random seeds, periods, stablecoins, and interaction categories would support the existence of a robust common scaling pattern. In contrast, systematic differences would suggest that the tail behaviour depends on temporal conditions, stablecoin-specific characteristics or interaction type.

\subsection{Composition weight analysis}
\label{subsec:composition_analysis}

To examine whether temporal variation in the overall scaling exponent can be reproduced by changes in the relative proportions of the four interaction categories, we constructed a composition-only counterfactual exponent. The overall scaling exponent is the exponent estimated from the pooled data composed of all transaction types. The analysis was conducted separately for both stablecoins: USDT and USDC at five fitting sample sizes, \(n \in
\{
50{,}000,\,
100{,}000,\,
200{,}000,\,
400{,}000,\,
800{,}000
\}\) using five independent random seeds for each sample size.

Let $N_i(t)$ denote the number of transactions belonging to interaction category $i$ during period $t$, where, \(i \in
\{
\mathrm{EOA\!-\!EOA},
\mathrm{EOA\!-\!SC},
\mathrm{SC\!-\!EOA},
\mathrm{SC\!-\!SC}
\}
\). The period-specific transaction weight of category $i$ was defined as

\begin{equation}
w_i(t)
=
\frac{N_i(t)}
{\displaystyle\sum_{j=1}^{K}N_j(t)},
\quad K=4,\quad \text{such that,}\quad \sum_{i=1}^{K}w_i(t)=1.
\label{eq:category_weight}
\end{equation}

Let $\hat{\alpha}_{i,t}^{(n,s)}$ denote the scaling exponent estimated for category $i$, period $t$, fitting sample size $n$, and random seed $s$. To remove temporal changes in the category-specific exponents, a fixed reference exponent was calculated for each category by averaging its estimated exponent across the $T=6$ study periods:

\begin{equation}
\alpha_{i,\mathrm{ref}}^{(n,s)}
=
\frac{1}{T}
\sum_{t=1}^{T}
\hat{\alpha}_{i,t}^{(n,s)}
\qquad T=6
\label{eq:reference_exponent}
\end{equation}

The reference exponent was calculated separately for each stablecoin, interaction category, fitting sample size, and random seed. Using these fixed reference exponents, the composition-only counterfactual exponent was constructed as

$$
\alpha_{\mathrm{comp},t}^{(n,s)}
=
\sum_{i=1}^{K}
w_i(t)\,
\alpha_{i,\mathrm{ref}}^{(n,s)}
\label{eq:composition_exponent}
$$

In this construction, the category-specific reference exponents remain constant across periods, while only the category weights $w_i(t)$ vary. Therefore, temporal variation in $\alpha_{\mathrm{comp},t}^{(n,s)}$ represents the variation generated by changes in transaction composition alone.

The counterfactual exponent was compared with the exponent fitted directly to the pooled transaction-value distribution, denoted by \(\displaystyle\alpha_{\mathrm{actual},t}^{(n,s)}.\) For graphical comparison, the actual and composition-only exponents were averaged independently across the $S=5$ random seeds. The seed-averaged exponent was calculated as

$$
\overline{\alpha}_{q,t}^{(n)}
=
\frac{1}{S}
\sum_{s=1}^{S}
\alpha_{q,t}^{(n,s)}
\qquad
q\in\{\mathrm{actual},\mathrm{comp}\}
\qquad S=5
\label{eq:seed_averaged_path}
$$

where $q$ identifies the fitted Overall or composition-only exponent series. The corresponding sample standard deviation across seeds was calculated as $ \sigma_{q,t}^{(n)} = \sqrt{\frac{1}{S-1}\sum_{s=1}^{S}\left[ \alpha_{q,t}^{(n,s)} - \overline{\alpha}_{q,t}^{(n)} \right]^2}$

The seed-averaged values, $\overline{\alpha}_{q,t}^{(n)}$, were used as the points in the temporal comparison, while $\sigma_{q,t}^{(n)}$ was used for the corresponding error bars.

To compare the magnitude of temporal variation in the two paths, the temporal range was calculated across the six periods. For the directly fitted Overall exponent, the range was defined as

$$
\Delta\alpha_{\mathrm{actual}}^{(n)}
=
\max_t
\left[
\overline{\alpha}_{\mathrm{actual},t}^{(n)}
\right]
-
\min_t
\left[
\overline{\alpha}_{\mathrm{actual},t}^{(n)}
\right]
\label{eq:actual_temporal_range}
$$

whereas the range of the composition-only exponent was defined as

$$
\Delta\alpha_{\mathrm{comp}}^{(n)}
=
\max_t
\left[
\overline{\alpha}_{\mathrm{comp},t}^{(n)}
\right]
-
\min_t
\left[
\overline{\alpha}_{\mathrm{comp},t}^{(n)}
\right]
\label{eq:composition_temporal_range}
$$

The relative counterfactual range was then calculated as

$$
C_{\mathrm{range}}^{(n)}
=
100
\frac{
\Delta\alpha_{\mathrm{comp}}^{(n)}
}{
\Delta\alpha_{\mathrm{actual}}^{(n)}
}
\label{eq:relative_composition_range}
$$

A small value of $C_{\mathrm{range}}^{(n)}$ indicates that changing category weights generates a substantially narrower temporal range than that observed in the exponent fitted directly to the pooled distribution. A value close to $100\%$ would indicate that the two paths have comparable temporal ranges. However, this measure compares only the magnitude of their ranges and does not imply that the two paths follow the same temporal direction.

\section{Data description}
\label{sec:data_description}

In this study, we used Ethereum blockchain transactions data for the stablecoins USDT and USDC from June 2024 to February 2026~\cite{zheng2020xblock,xblockwebsite}. The timestamps in Table~\ref{tab:study_periods} indicate an observation window from 1 June 2024 to 20 February 2026. Before performing the analysis, we preprocessed the data. The Ethereum blockchain supports thousands of tokens, including many stablecoins. Here, we focused only on USDT and USDC, as they represent the two largest stablecoins by market capitalization and together account for around 80\% of the stablecoin market. We identified these stablecoins using their official contract addresses obtained from Etherscan~\cite{etherscan}. The USDT contract address is "0xdac17f958d2ee523a2206206994597c13d831ec7", and the USDC contract address is "0xa0b86991c6218b36c1d19d4a2e9eb0ce3606eb48". For the temporal analysis, the transaction data were divided into six study periods, with each period combining approximately three consecutive transaction-data batches. Table~\ref{tab:study_periods} reports the initial and final Unix timestamps and their corresponding UTC dates and times. Table~\ref{tab:dataset_description} summarizes the variables used in this study.

\par\medskip

\noindent
{\centering

\begin{tabular}{ccrrll}
\hline
Study period  &
Initial timestamp & Final timestamp &
Initial date and time (UTC) & Final date and time (UTC) \\
\hline

Period 1  &
1717281407 & 1726330811 &
2024-06-01 22:36:47 & 2024-09-14 16:20:11 \\

Period 2 &
1726330823 & 1735377275 &
2024-09-14 16:20:23 & 2024-12-28 09:14:35 \\

Period 3  &
1735377287 & 1744426259 &
2024-12-28 09:14:47 & 2025-04-12 02:50:59 \\

Period 4  &
1744426271 & 1753492919 &
2025-04-12 02:51:11 & 2025-07-26 01:21:59 \\

Period 5  &
1753492931 & 1762550015 &
2025-07-26 01:22:11 & 2025-11-07 21:13:35 \\

Period 6 &
1762550027 & 1771613831 &
2025-11-07 21:13:47 & 2026-02-20 18:57:11 \\
\hline
\end{tabular}

\par}

\refstepcounter{table}
\label{tab:study_periods}

\par\smallskip

\noindent
\textbf{Table \ref{tab:study_periods}.}
The table contains the definitions of the six study periods used in the temporal analysis. Each period combines three consecutive transaction-data batches. The initial and final dates and times were obtained by converting the corresponding Unix timestamps to Coordinated Universal Time (UTC).

\par\medskip

In the Ethereum blockchain, two types of accounts exist: Externally Owned Accounts (EOAs), which are controlled by users through public-private cryptographic keys, and Smart Contracts (SCs), which are controlled by executable program code~\cite{chen2020traveling,kolvart2016smart,szabo1997idea}. Before the analysis, we removed the missing values from the data. The fromIsContract and toIsContract indicators provided in the dataset were used to identify whether the sender and receiver were EOAs or SCs. Based on the interaction between the sender and receiver, we further classified the stablecoin transactions into four categories: EOA-EOA, EOA-SC, SC-EOA, and SC-SC. Table~\ref{tab:stablecoin_transactions} presents the distribution of transaction counts across these four categories. The time is stored as a Unix timestamp. For example, the timestamp 1767225600 corresponds to 2026-01-01 00:00:00 UTC under the Unix epoch system. This time is begins at 1970-01-01 00:00:00 UTC~\cite{unix_time_wikipedia}. For temporal analysis and easier interpretation, we converted all timestamps into the human-readable format YYYY-MM-DD HH:MM:SS in UTC.

\par\medskip

\noindent
{\centering

\renewcommand{\arraystretch}{1.2}

\begin{tabular}{p{3.3cm} p{10.5cm}}
\hline
\textbf{Column} & \textbf{Description} \\
\hline

\textbf{timeStamp}
& Unix timestamp indicating when the block containing the transaction was confirmed. \\

\textbf{tokenAddress}
& Contract address that uniquely identifies a stablecoin on the Ethereum blockchain. \\

\textbf{fromIsContract}
& Indicates whether the sender is a smart contract (SC = 1) or an externally owned account (EOA = 0). \\

\textbf{toIsContract}
& Indicates whether the receiver is a smart contract (SC = 1) or an externally owned account (EOA = 0). \\

\textbf{value}
& Raw token transfer amount recorded in the smallest divisible unit of USDT or USDC. \\

\hline
\end{tabular}

\par}

\refstepcounter{table}
\label{tab:dataset_description}

\par\smallskip

\noindent
\textbf{Table \ref{tab:dataset_description}.}
The table contains the description of the stablecoin transaction dataset and the variables used in the analysis. The dataset includes transaction times, token contract addresses, account-type indicators, and token transfer amounts.

\par\medskip

The stablecoin transfer amount is stored in the smallest divisible token unit rather than directly in its standard decimal denomination. Both USDT and USDC use six decimal places, meaning that the raw transfer amount is scaled by \(10^6\). Therefore, the nominal USD-denominated transaction amount was obtained using

\begin{equation}
\mathrm{Nominal\ USD\ amount}
=
\frac{\mathrm{value}}{1000000}
\label{eq:usd_conversion}
\end{equation}

For example, a raw value of \(1000000\) corresponds to one USDT or one USDC and therefore represents a nominal USD-denominated transaction amount of 1. This conversion changes the raw token units into their standard decimal denomination; it does not independently measure the historical market price of the stablecoin at the time of transfer.

We used the preprocessed USD value transaction amounts in all subsequent analyses to investigate statistical scaling in the USDT and USDC transaction-value distributions.

\par\medskip

\noindent
{\centering

\renewcommand{\arraystretch}{1.5}

\resizebox{\textwidth}{!}{
\begin{tabular}{llrrrrr}
\hline
\textbf{Period} & \textbf{Stablecoin} &
\textbf{EOA--EOA} & \textbf{EOA--SC} &
\textbf{SC--EOA} & \textbf{SC--SC} &
\textbf{Overall Transactions} \\
\hline

Period 1 & USDT & 10,805,193 & 1,421,431 & 2,142,502 & 3,097,961 & 17,467,087 \\
         & USDC & 2,966,675 & 1,172,403 & 1,903,353 & 3,123,789 & 9,166,220 \\

Period 2 & USDT & 11,219,446 & 1,611,866 & 2,542,764 & 2,815,891 & 18,189,967 \\
         & USDC & 3,433,497 & 1,439,055 & 2,271,114 & 3,132,574 & 10,276,240 \\

Period 3 & USDT & 14,102,556 & 1,656,001 & 3,128,174 & 5,090,133 & 23,976,864 \\
         & USDC & 6,505,209 & 2,070,672 & 3,700,506 & 6,599,879 & 18,876,266 \\

Period 4 & USDT & 18,045,735 & 1,814,107 & 3,886,928 & 5,937,174 & 29,683,944 \\
         & USDC & 10,372,970 & 2,747,044 & 4,503,959 & 8,008,127 & 25,632,100 \\

Period 5 & USDT & 23,859,906 & 2,664,861 & 6,140,180 & 9,567,876 & 42,232,823 \\
         & USDC & 12,622,874 & 4,339,690 & 7,109,532 & 13,867,821 & 37,939,917 \\

Period 6 & USDT & 48,926,380 & 3,125,113 & 13,903,739 & 13,156,221 & 79,111,453 \\
         & USDC & 25,190,871 & 4,504,301 & 11,314,628 & 16,044,850 & 57,054,650 \\

\hline
\textbf{Total} & \textbf{USDT} &
\textbf{126,959,216} &
\textbf{12,293,379} &
\textbf{31,744,287} &
\textbf{39,665,256} &
\textbf{210,662,138} \\

& \textbf{USDC} &
\textbf{61,092,096} &
\textbf{16,273,165} &
\textbf{30,803,092} &
\textbf{50,776,040} &
\textbf{158,945,393} \\
\hline
\end{tabular}
}

\par}

\refstepcounter{table}
\label{tab:stablecoin_transactions}

\par\smallskip

\noindent
\textbf{Table \ref{tab:stablecoin_transactions}.}
The table contains the number of USDT and USDC transactions across the four interaction categories and six study periods. The dataset comprises 210,662,138 USDT transactions and 158,945,393 USDC transactions, giving a total of 369,607,531 transactions.

\par\medskip

\section{Result}
\label{sec:Result}

In this section, we present the results of our analysis based on the normalized USD transaction values. We first examine whether stablecoin transaction-value tails are compatible with power-law scaling across the analysed categories and periods. We then examine the robustness of the estimated exponents through sensitivity analysis. Finally, we perform a counterfactual analysis to test whether changes in category weights can explain the temporal variation in the overall exponent.

\subsection{Power-Law scaling and interaction-specific tail exponents}
\label{subsec:scaling_exponents}

\par\medskip

\noindent
{\centering

\begin{minipage}{0.485\linewidth}
    \centering
    \includegraphics[width=\linewidth]{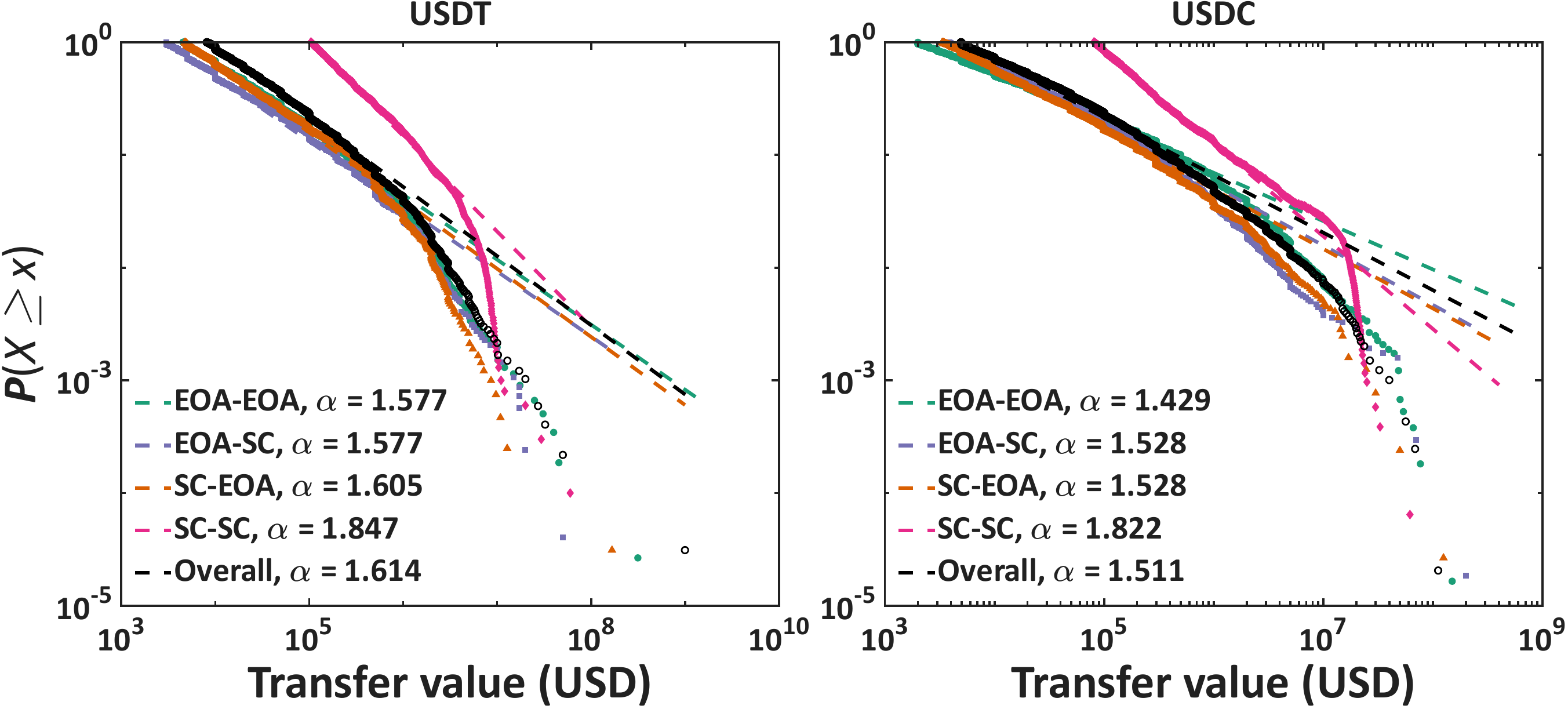}

    \small (a) Period 1
\end{minipage}
\hfill
\begin{minipage}{0.485\linewidth}
    \centering
    \includegraphics[width=\linewidth]{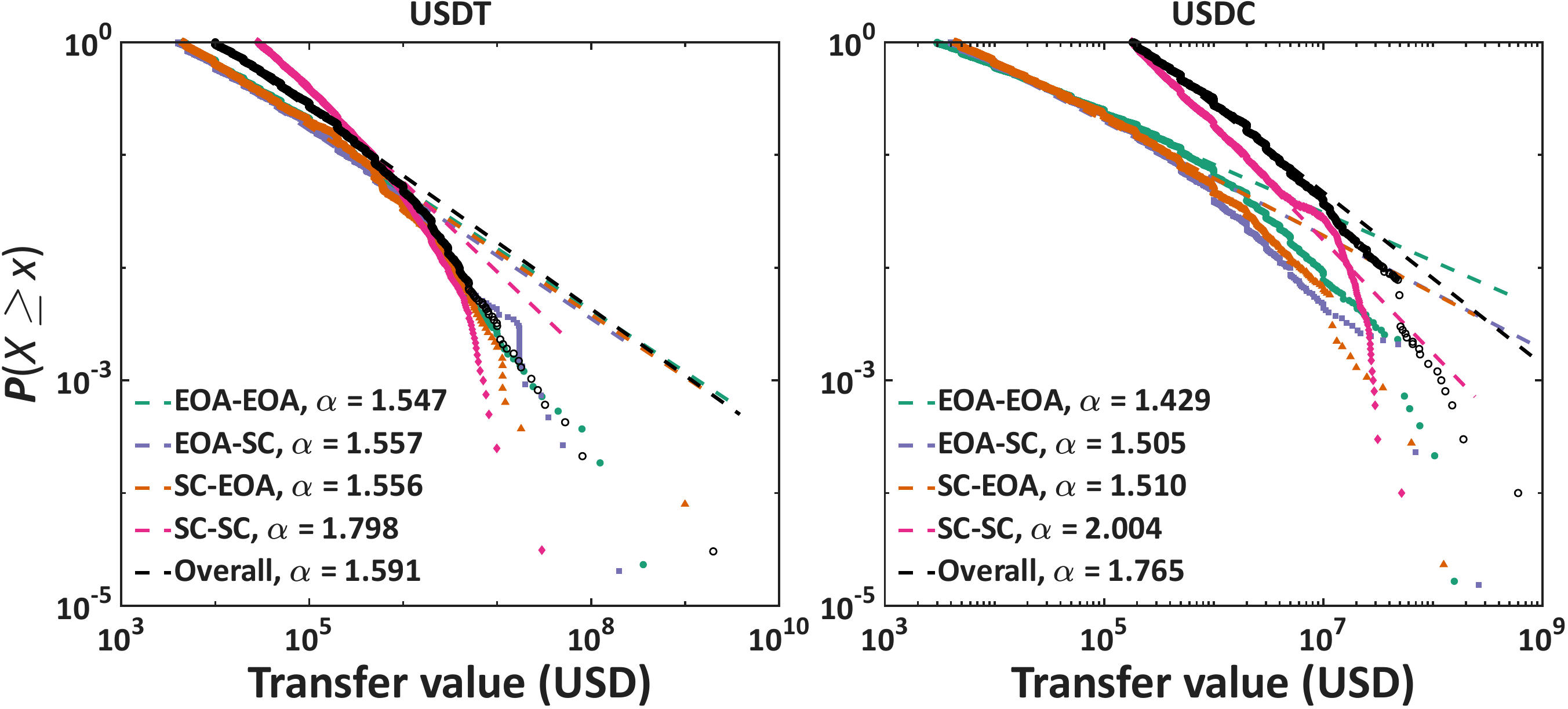}

    \small (b) Period 2
\end{minipage}

\vspace{-0.05cm}

\begin{minipage}{0.485\linewidth}
    \centering
    \includegraphics[width=\linewidth]{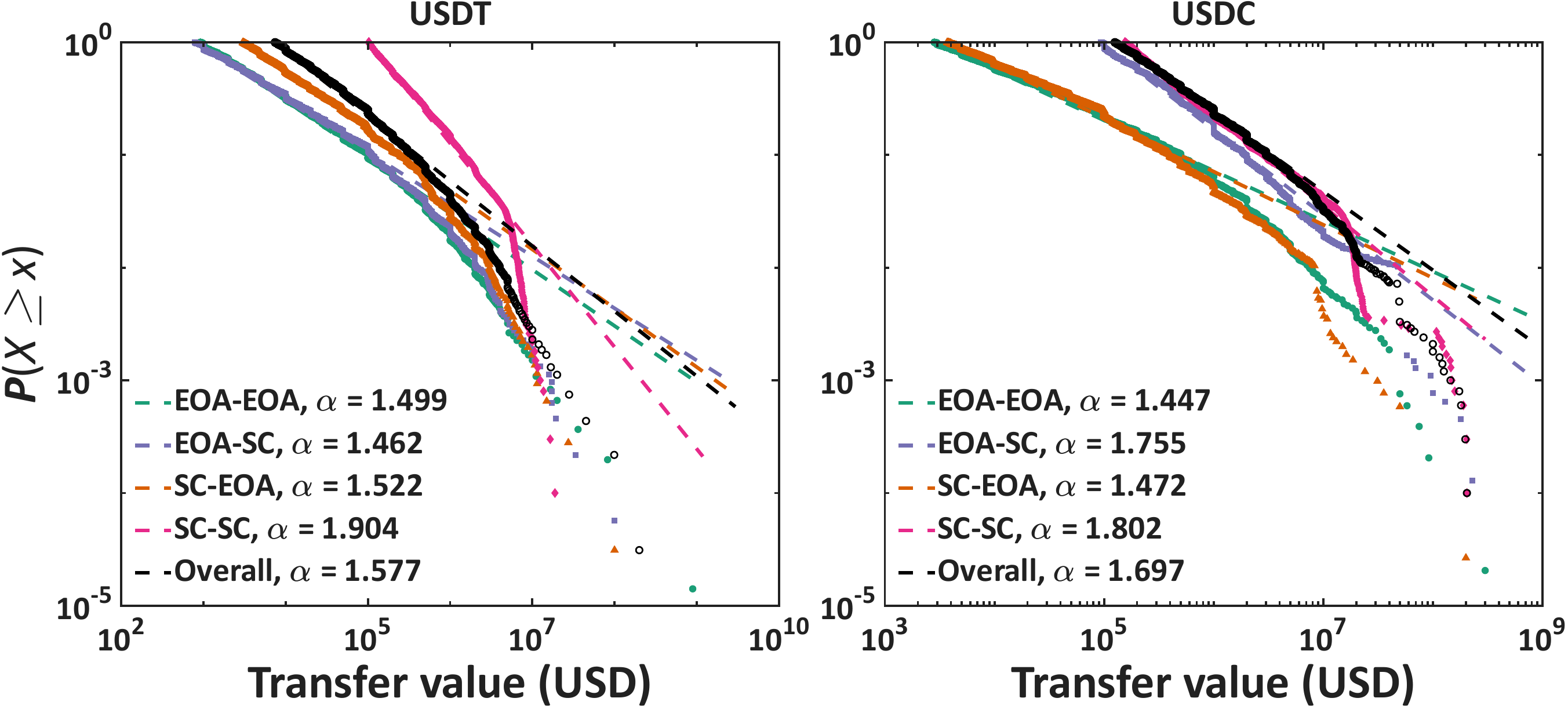}

    \small (c) Period 3
\end{minipage}
\hfill
\begin{minipage}{0.485\linewidth}
    \centering
    \includegraphics[width=\linewidth]{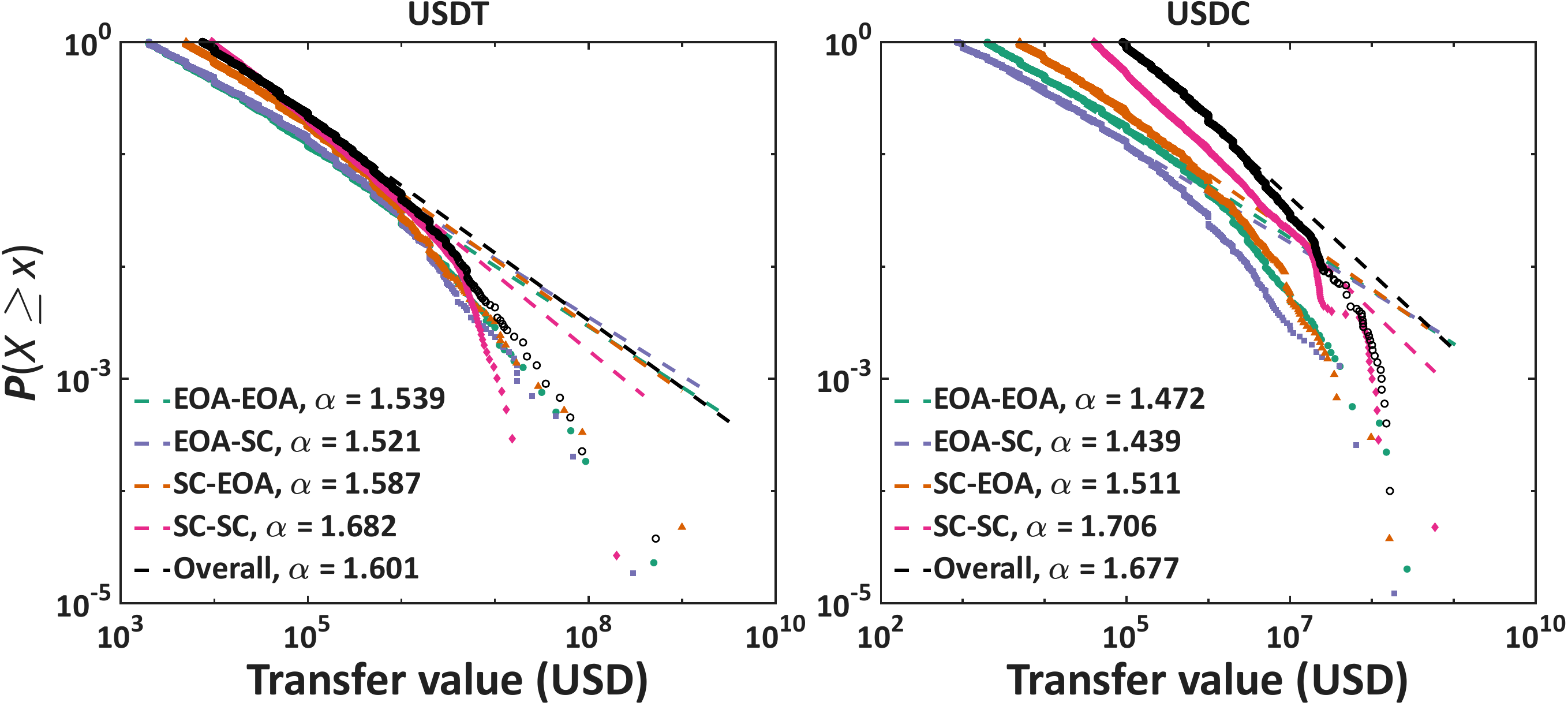}

    \small (d) Period 4
\end{minipage}

\vspace{-0.05cm}

\begin{minipage}{0.485\linewidth}
    \centering
    \includegraphics[width=\linewidth]{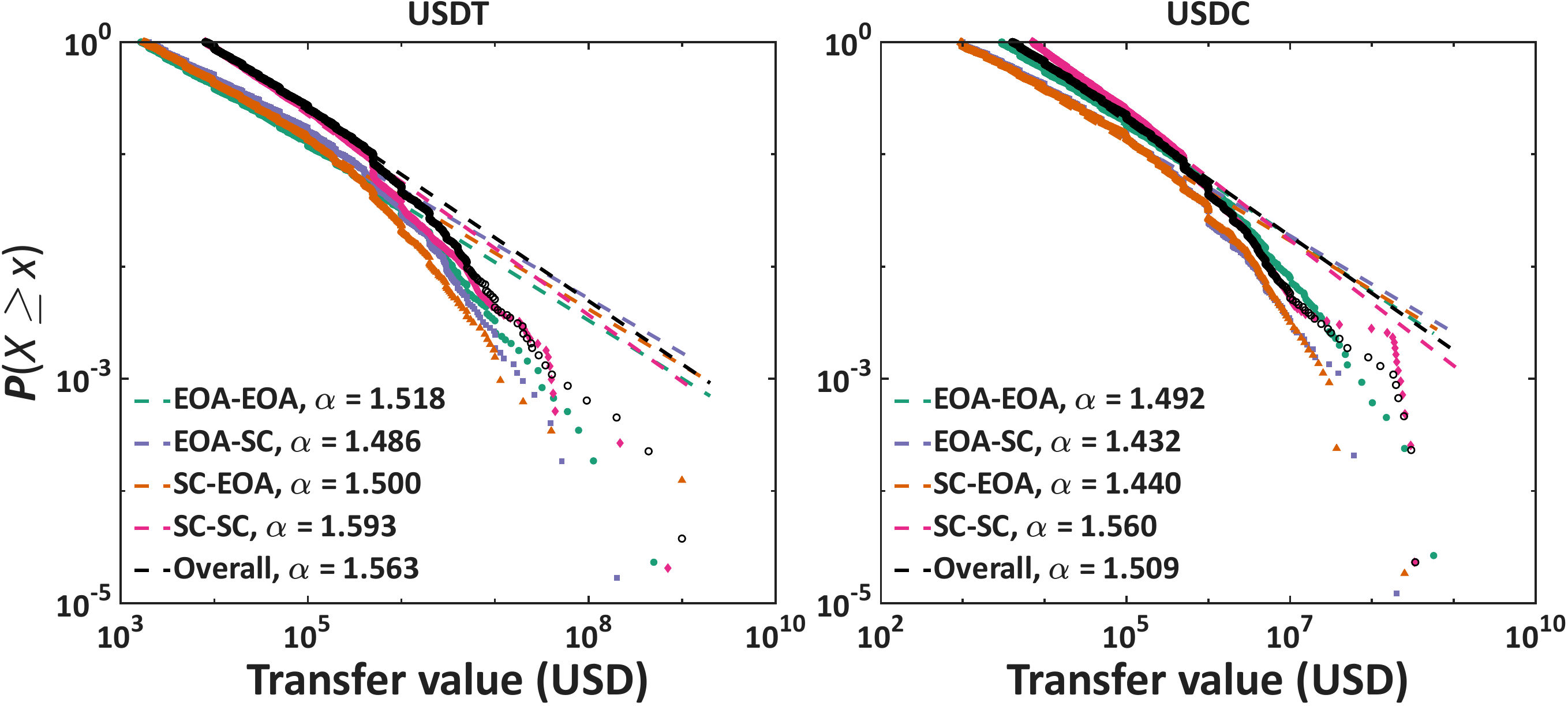}

    \small (e) Period 5
\end{minipage}
\hfill
\begin{minipage}{0.485\linewidth}
    \centering
    \includegraphics[width=\linewidth]{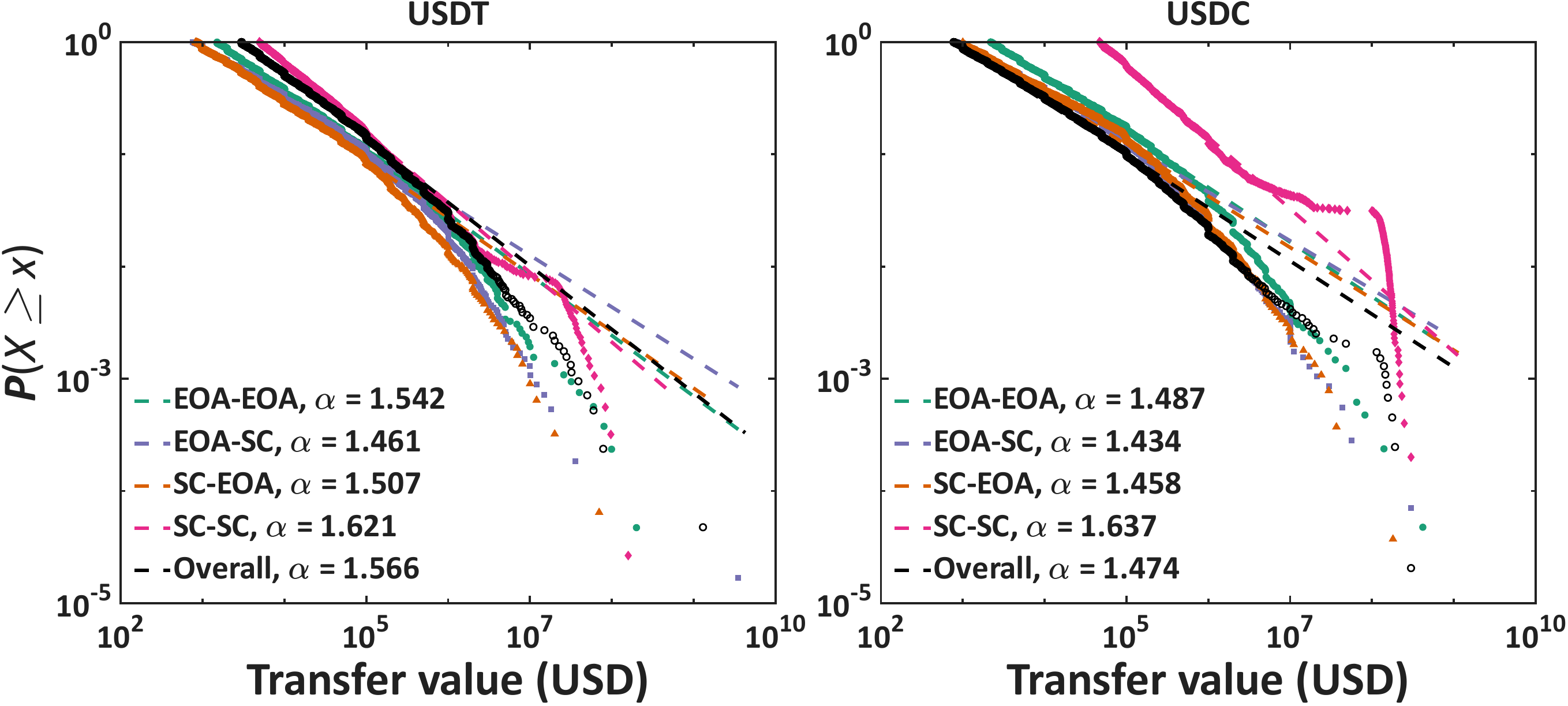}

    \small (f) Period 6
\end{minipage}

\par}

\refstepcounter{figure}
\label{fig:ccdf_all_periods}

\par\smallskip

\noindent
\textbf{Fig. \ref{fig:ccdf_all_periods}.}
The figure represents the complementary cumulative distribution functions (CCDFs) of stablecoin transaction values across six periods. For each period, the left panel shows USDT and the right panel shows USDC. The empirical transaction-value distributions and the corresponding maximum-likelihood power-law fits above the lower cut-off, $x_{\min}$, are shown for the overall, EOA-EOA, EOA-SC, SC-EOA, and SC-SC interaction categories. The estimated scaling exponents, $\alpha$, are reported in the legends. The power-law form is observed across all interaction categories, while the SC-SC category exhibits larger scaling exponents than the EOA-involved categories.

\par\medskip

Scaling and universality are fundamental concepts in the statistical physics of financial markets. Scaling behavior is often characterized by power-law tails in distributions, while universality refers to the approximate invariance of scaling exponents across different assets, markets, or time periods. Such scaling behaviour and approximate universality have been reported for quantities such as stock returns, trade sizes, trading volume, intertrade times, and price impact across diverse markets~\cite{stanley2008statistical,beguvsic2018scaling,plerou2008stock,gopikrishnan1998inverse,sato2025strict}. In the context of stablecoin transactions, however, it remains unclear whether such scaling behaviour is universal or depends on the underlying interaction structure. Unlike traditional market data, stablecoin transfers on the Ethereum blockchain provide a unique high-resolution dataset for testing this universality. Here, we estimated power-law exponents for USDT and USDC across the six periods and four interaction categories: EOA-EOA, EOA-SC, SC-EOA, and SC-SC, along with an overall category.

Figure~\ref{fig:ccdf_all_periods} shows the complementary cumulative distribution functions (CCDFs) of stablecoin transaction values for USDT and USDC across the six periods, estimated using a fitting sample size of $N = 200{,}000$. The solid lines in the figure~\ref{fig:ccdf_all_periods} represent the empirical CCDFs, while the dashed lines show the power-law fits obtained using the maximum-likelihood method for data above the lower cut-off, $x_{\min}$. The $x_{\min}$ was chosen by minimizing the Kolmogorov-Smirnov (KS) distance between the empirical and fitted cumulative distributions. The corresponding scaling exponent ($\alpha$) was then calculated using maximum likelihood. The resulting $\alpha$ for each category is given in the legend. The empirical distributions for both stablecoins (USDT and USDC) and all interaction categories are well described by the power-law form above  $x_{\min}$.

To summarize the temporal behavior, we calculate the average exponent $\bar{\alpha}_i = T^{-1}\sum_{t=1}^{T} \hat{\alpha}_{i,t}$ for each category $i$ across the $T=6$ study periods, with the uncertainty reported as the standard deviation across periods. For the overall category, the average exponent is $\bar{\alpha} = 1.566 \pm 0.019$ for USDT and $\bar{\alpha} = 1.605 \pm 0.104$ for USDC. Among the interaction categories, EOA-EOA gives $\bar{\alpha} = 1.542 \pm 0.023$ for USDT and $\bar{\alpha} = 1.459 \pm 0.025$ for USDC. The EOA--SC category gives $\bar{\alpha} = 1.521 \pm 0.040$ for USDT and $\bar{\alpha} = 1.484 \pm 0.050$ for USDC, while the SC-EOA category gives $\bar{\alpha} = 1.544 \pm 0.043$ for USDT and $\bar{\alpha} = 1.486 \pm 0.033$ for USDC. The SC-SC interaction category shows the largest average exponent among the four: $\bar{\alpha} = 1.727 \pm 0.120$ for USDT and $\bar{\alpha} = 1.722 \pm 0.148$ for USDC. For all interaction categories, the scaling exponents fall in the range $1.4 \lesssim \alpha \lesssim 1.8$, confirming heavy-tailed behavior across all periods and for both stablecoins.  The three categories involving EOAs interaction cluster between $\alpha \approx 1.45$--$1.60$, whereas SC-SC interactions exhibit slightly larger exponents around $\alpha \approx 1.72$--$1.73$. This separation, observed across all periods and for both stablecoins, indicates that the scaling exponent depends on whether transactions involve EOAs or SCs. The power-law form holds across all categories, periods, and stablecoins, and the scaling exponent in the range $1.4 \lesssim \alpha \lesssim 1.8$ is consistent with scaling behavior previously reported in traditional financial markets. These results show that heavy-tailed, power-law behaviour is consistently observed across the analysed samples.

\subsection{Sensitivity analysis of the scaling exponents}
\label{subsec:Sensitivity Analysis}

\begin{table}[t!]
\centering
\caption{Table contains the sensitivity analysis of the estimated power-law exponent, $\alpha$, with respect to the fitting sample size. For each interaction category and sample size $N$, the reported value is the mean $\pm$ standard deviation across the six study periods. The mean exponents remain stable as the fitting sample size increases from $50{,}000$ to $800{,}000$, indicating limited sensitivity to the choice of $N$.}
\label{tab:sensitivity}

\vspace{4pt}
\textbf{USDT}\\[2pt]
\resizebox{\textwidth}{!}{
\begin{tabular}{lccccc}
\toprule
Interaction & $N=50{,}000$ & $N=100{,}000$ & $N=200{,}000$ & $N=400{,}000$ & $N=800{,}000$ \\
\midrule
Overall  & $1.581 \pm 0.028$ & $1.589 \pm 0.026$ & $1.585 \pm 0.023$ & $1.586 \pm 0.023$ & $1.588 \pm 0.021$ \\
EOA--EOA & $1.534 \pm 0.023$ & $1.544 \pm 0.015$ & $1.539 \pm 0.025$ & $1.544 \pm 0.025$ & $1.544 \pm 0.021$ \\
EOA--SC  & $1.511 \pm 0.041$ & $1.508 \pm 0.046$ & $1.509 \pm 0.048$ & $1.510 \pm 0.049$ & $1.507 \pm 0.049$ \\
SC--EOA  & $1.539 \pm 0.035$ & $1.545 \pm 0.036$ & $1.541 \pm 0.041$ & $1.539 \pm 0.040$ & $1.533 \pm 0.040$ \\
SC--SC   & $1.739 \pm 0.128$ & $1.741 \pm 0.127$ & $1.747 \pm 0.132$ & $1.740 \pm 0.127$ & $1.740 \pm 0.127$ \\
\bottomrule
\end{tabular}}

\vspace{8pt}
\textbf{USDC}\\[2pt]
\resizebox{\textwidth}{!}{
\begin{tabular}{lccccc}
\toprule
Interaction & $N=50{,}000$ & $N=100{,}000$ & $N=200{,}000$ & $N=400{,}000$ & $N=800{,}000$ \\
\midrule
Overall  & $1.583 \pm 0.090$ & $1.601 \pm 0.104$ & $1.596 \pm 0.117$ & $1.597 \pm 0.115$ & $1.590 \pm 0.110$ \\
EOA--EOA & $1.459 \pm 0.030$ & $1.456 \pm 0.028$ & $1.459 \pm 0.029$ & $1.458 \pm 0.029$ & $1.458 \pm 0.029$ \\
EOA--SC  & $1.526 \pm 0.072$ & $1.546 \pm 0.100$ & $1.555 \pm 0.115$ & $1.523 \pm 0.118$ & $1.535 \pm 0.108$ \\
SC--EOA  & $1.482 \pm 0.044$ & $1.478 \pm 0.043$ & $1.494 \pm 0.040$ & $1.487 \pm 0.036$ & $1.481 \pm 0.034$ \\
SC--SC   & $1.735 \pm 0.168$ & $1.736 \pm 0.179$ & $1.749 \pm 0.163$ & $1.749 \pm 0.164$ & $1.753 \pm 0.160$ \\
\bottomrule
\end{tabular}}
\end{table}

To assess the robustness of the estimated scaling exponents and the evidence for the robustness of the observed scaling patterns in USDT and USDC blockchain transaction values, we perform a sensitivity analysis by varying different fitting sample sizes and random seeds. We repeat the exponent estimation for five fitting sample sizes, $N \in \{50{,}000,\ 100{,}000,\ 200{,}000,\ 400{,}000,\ 800{,}000\}$, and five independent random seeds for each sample size. The analysis is conducted separately for the overall, EOA-EOA, EOA-SC, SC-EOA, and SC-SC interaction categories. For every combination of fitting sample size and random seed, the lower cut-off, $x_{\min}$, and scaling exponent, $\alpha$, are re-estimated following the procedure described in Section~\ref{sec:Method_Sensitivity}. For each fitting sample size, the mean exponent and its standard deviation are calculated across the five random seeds and presented in Figure~\ref{fig:alpha_sample_size_sensitivity}.

\par\medskip

\noindent
{\centering

% Border thickness
\setlength{\fboxrule}{0.6pt}

% Space between the image and border
\setlength{\fboxsep}{1.5pt}

% =========================================================
% ROW 1: 50k and 200k
% =========================================================

\begin{minipage}[t]{0.485\textwidth}
    \centering
    \includegraphics[
        width=\dimexpr\linewidth-2\fboxsep-2\fboxrule\relax
    ]{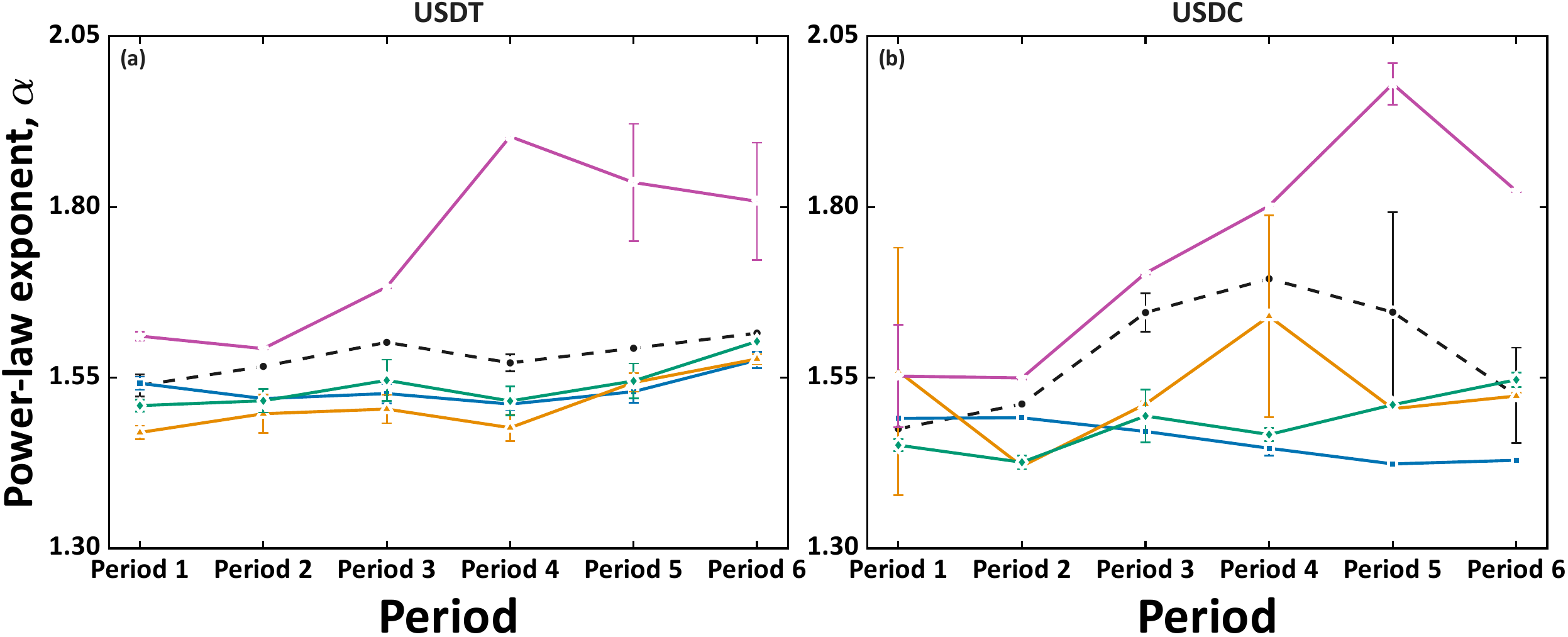}

    \small (a) N $=50{,}000$
    \refstepcounter{subfigure}
    \label{fig:alpha_50k}
\end{minipage}
\hfill
\begin{minipage}[t]{0.485\textwidth}
    \centering
    \includegraphics[
        width=\dimexpr\linewidth-2\fboxsep-2\fboxrule\relax
    ]{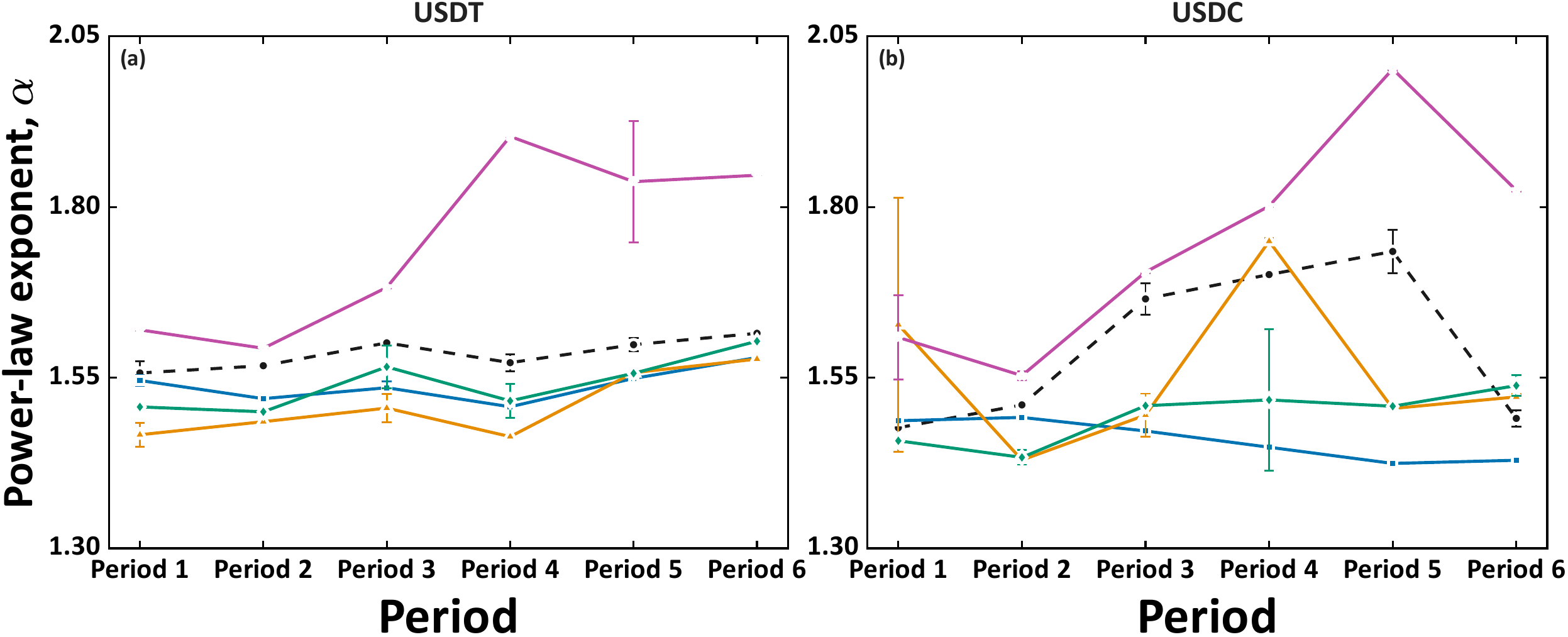}

    \small (b) N$=200{,}000$
    \refstepcounter{subfigure}
    \label{fig:alpha_100k}
\end{minipage}

\vspace{0.35cm}

% =========================================================
% ROW 2: 400k and 800k
% =========================================================

\begin{minipage}[t]{0.485\textwidth}
    \centering
    \includegraphics[
        width=\dimexpr\linewidth-2\fboxsep-2\fboxrule\relax
    ]{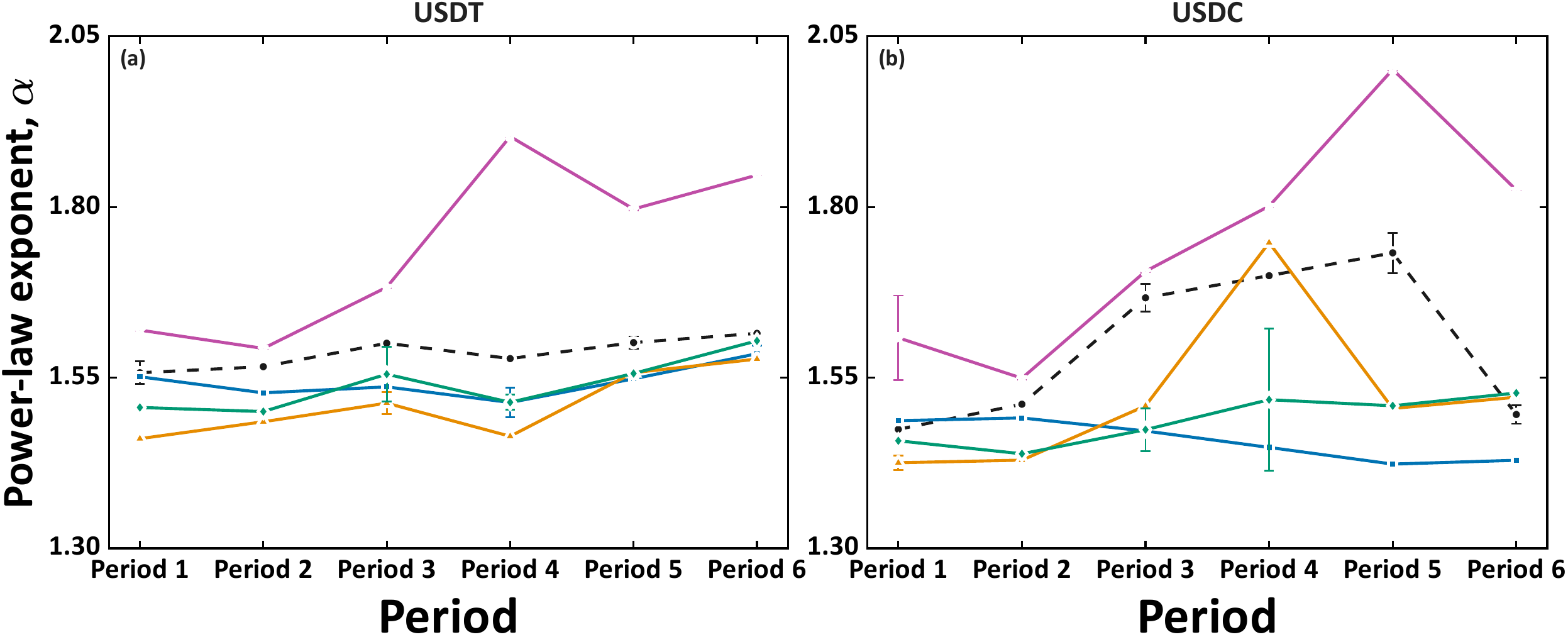}

    \small (c) N$=400{,}000$
    \refstepcounter{subfigure}
    \label{fig:alpha_200k}
\end{minipage}
\hfill
\begin{minipage}[t]{0.485\textwidth}
    \centering
    \includegraphics[
        width=\dimexpr\linewidth-2\fboxsep-2\fboxrule\relax
    ]{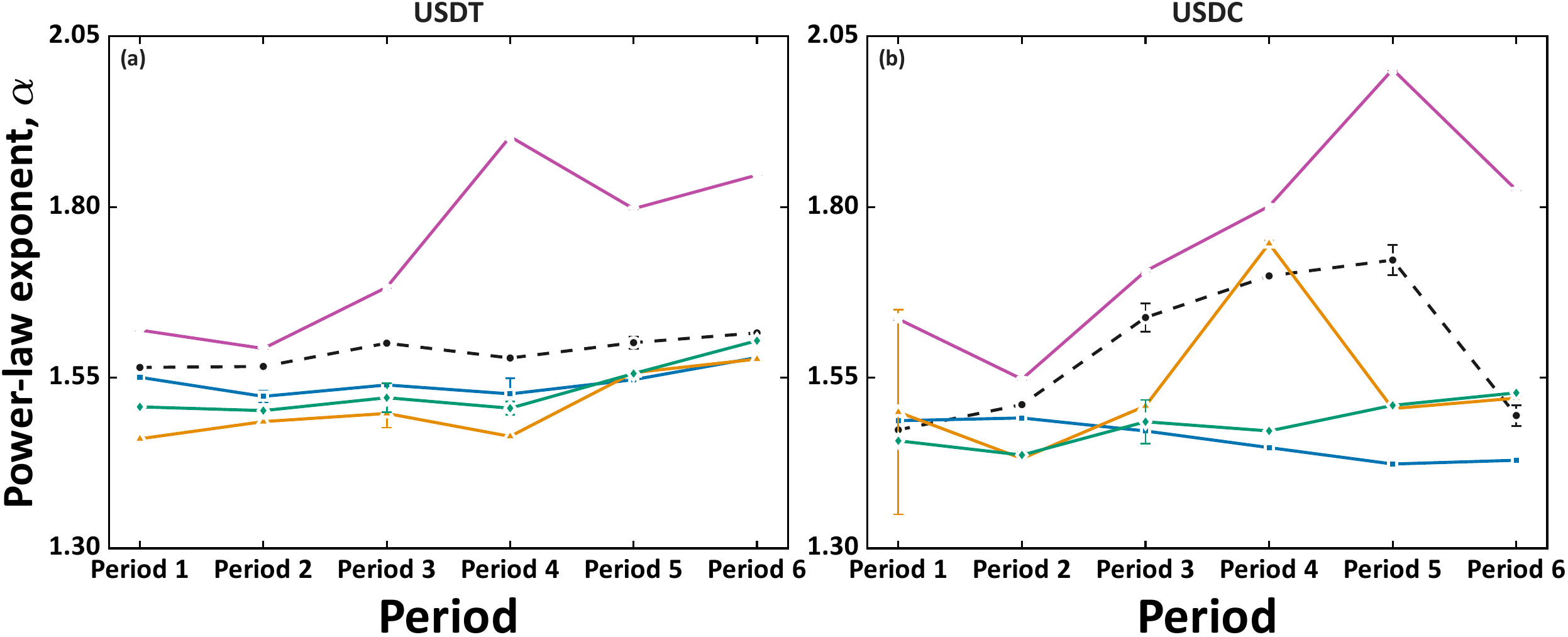}

    \small (d) N$=800{,}000$
    \refstepcounter{subfigure}
    \label{fig:alpha_400k}
\end{minipage}

\vspace{0.35cm}

% =========================================================
% ROW 3: 100k centred
% =========================================================

\begin{minipage}[t]{0.485\textwidth}
    \centering
    \includegraphics[
        width=\dimexpr\linewidth-2\fboxsep-2\fboxrule\relax
    ]{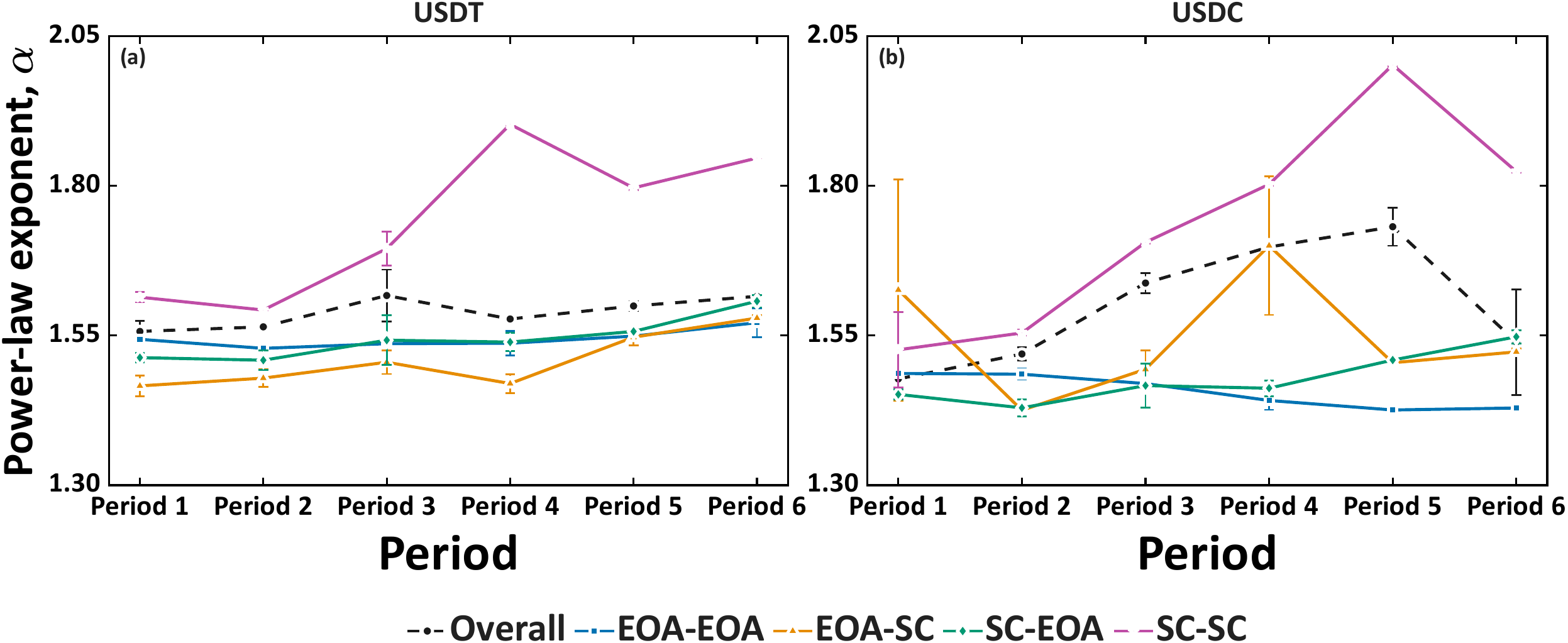}

    \small (e) N$=100{,}000$
    \refstepcounter{subfigure}
    \label{fig:alpha_800k}
\end{minipage}

\par}

\refstepcounter{figure}
\label{fig:alpha_sample_size_sensitivity}

\par\smallskip

\noindent
\textbf{Fig. \ref{fig:alpha_sample_size_sensitivity}.}
The figure represents the sensitivity of the temporal evolution of the interaction-specific power-law exponents to the fitting sample size. Each panel corresponds to a different fitting sample size, $N$. The markers denote the mean scaling exponent estimated across five independent random seeds, and the vertical error bars represent one standard deviation. The temporal patterns and interaction-specific differences remain consistent across all fitting sample sizes.

\par\medskip

Figure~\ref{fig:alpha_sample_size_sensitivity} shows the temporal evolution of the estimated scaling exponent, $\alpha$, across the six periods and five fitting sample sizes. Each sample-size panel presents USDT and USDC side by side. The coloured solid lines represent the mean exponents of the EOA-EOA, EOA-SC, SC-EOA, and SC-SC categories, while the dashed line represents the mean exponents of the overall category. Each marker denotes the mean exponent across five independent random seeds, and the vertical error bars indicate one standard deviation. As shown in Figure~\ref{fig:alpha_sample_size_sensitivity}, the interaction-specific and temporal patterns remain consistent across the five fitting sample sizes. The SC-SC category exhibits the largest exponents and relatively higher temporal variation, whereas the three EOA-involved categories remain concentrated at lower values. The Overall exponent falls between the lower values seen for the EOA-involved categories and the higher value for SC–SC, consistent with the earlier results. This intermediate position of the overall category reflects the pooled nature, which combines transactions from all four interaction types, i.e EOAs and SC. The persistence of the same interaction-specific structure across all fitting sample sizes provides further evidence that the observed scaling behaviour is not an artefact of sample selection.

Table~\ref{tab:sensitivity} shows how stable the estimated exponents are across different fitting sample sizes. For each sample size $N$, the reported mean $\bar{\alpha}(N)$ and standard deviation $\sigma_{\alpha}(N)$ are calculated in two steps. First, for each period, the exponent is averaged over the five random seeds; then, the mean and standard deviation are calculated across the six periods. The standard deviation therefore reflects temporal variation across periods rather than seed-to-seed uncertainty. These calculations are performed separately for each stablecoin, interaction category, and fitting sample size. Across both stablecoins and all interaction categories, the mean scaling exponents, as presented in the table~\ref{tab:sensitivity}, remain stable over the five fitting sample sizes. Increasing $N$, from $50{,}000$ to $800{,}000$, produces only small changes in the estimated mean exponents and does not alter the main temporal patterns. The variation associated with the choice of fitting sample size is generally smaller than the observed period-to-period variation. Moreover, the seed-to-seed uncertainty generally decreases at larger sample sizes, as indicated by the narrowing error bars in Figure~\ref{fig:alpha_sample_size_sensitivity}. These results show that the estimated scaling exponents are not sensitive to a particular fitting sample size, thereby supporting the robustness of the estimated exponents to fitting-sample selection in stablecoin transaction values reported in the above section~\ref{subsec:scaling_exponents}.

\subsection{Transaction composition and exponent variation}
\label{subsec:composition_results}

\par\medskip

\noindent
{\centering

\includegraphics[width=0.96\textwidth]{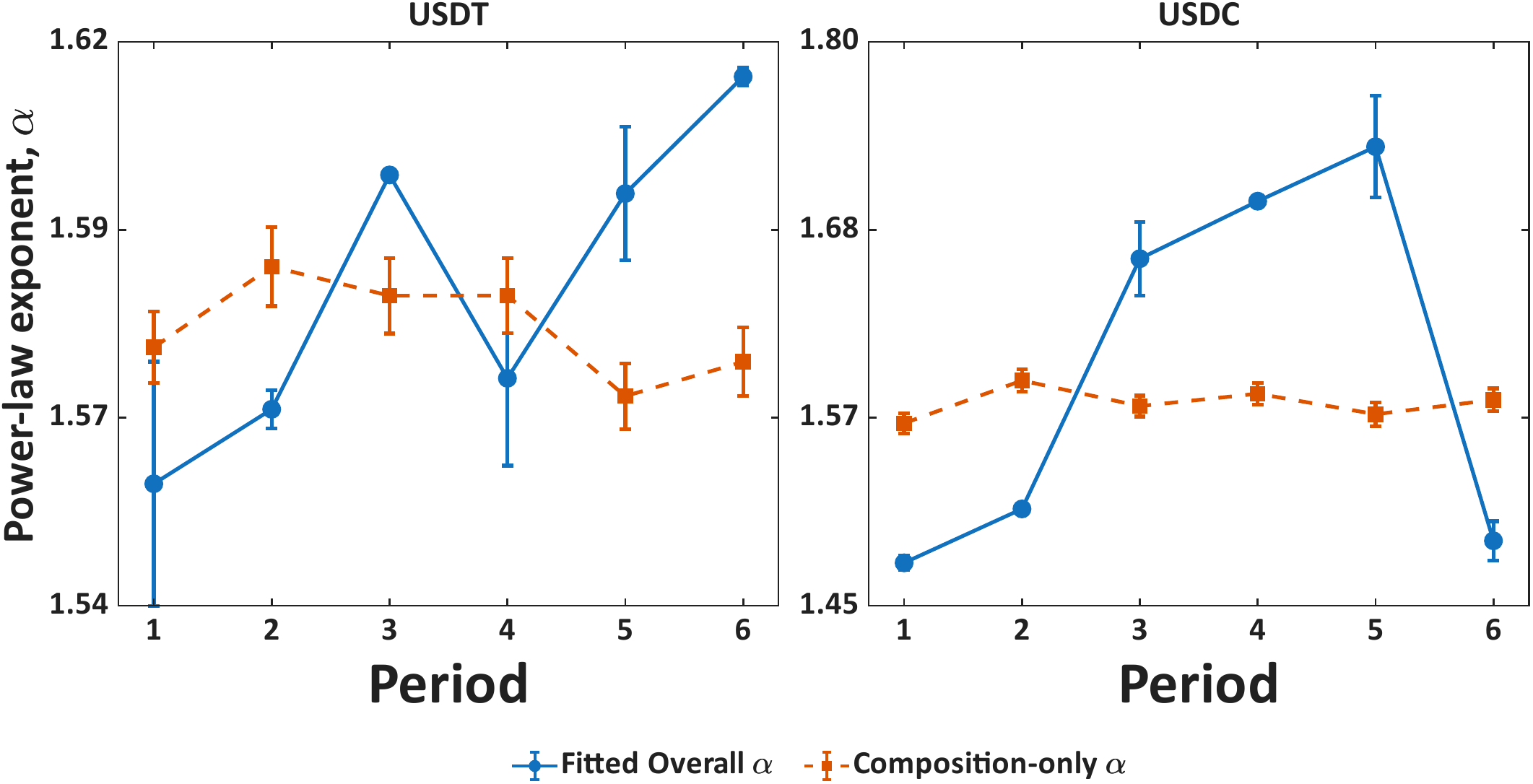}

\par}

\refstepcounter{figure}
\label{fig:actual_vs_composition}

\par\smallskip

\noindent
\textbf{Fig. \ref{fig:actual_vs_composition}.}
The figure presents the temporal evolution of the directly fitted overall and composition-only counterfactual exponents for USDT and USDC at $N=200{,}000$. The points represent means across five random seeds, and error bars denote $\pm1$ standard deviation. The weaker variation of the counterfactual exponent indicates that changes in transaction composition alone do not reproduce the observed overall exponent path.

\par\medskip

To examine whether changes in transaction weights drive temporal variation in the overall exponent, i.e., the exponent estimated from the pooled data of all transaction types, we compare the directly fitted overall exponent, $\alpha_{\mathrm{actual}}$, with a composition-only counterfactual exponent, $\alpha_{\mathrm{comp}}$. In the counterfactual calculation, we hold the reference exponent $\bar{\alpha}_i$ of each interaction category $i$ fixed at its six-period average, while allowing only its transaction weight $w_i(t)$ to vary across periods. The counterfactual exponent is constructed as $\alpha_{\mathrm{comp}}(t) = \sum_i w_i(t) \bar{\alpha}_i$. This $\alpha_{\mathrm{comp}}(t)$ represents what the overall exponent would be if only the category weights changed over time.

In Figure~\ref{fig:actual_vs_composition}, we compare $\alpha_{\mathrm{actual}}$ and $\alpha_{\mathrm{comp}}$ at the representative sample size $N=200{,}000$. Each point in the figure~\ref{fig:actual_vs_composition} represents the mean over five random seeds, with error bars showing $\pm1$ standard deviation. The figure shows that the composition-only paths are flatter than the fitted overall paths for both stablecoins. For USDT, the temporal range of $\alpha_{\mathrm{actual}}$ is $0.058$, compared with $0.018$ for $\alpha_{\mathrm{comp}}$. For USDC, the corresponding ranges are $\Delta\alpha_{\mathrm{actual}} = 0.258$ and $\Delta\alpha_{\mathrm{comp}} = 0.027$. Thus, the counterfactual exponent exhibits only a fraction of the temporal variation observed in the directly fitted overall exponent and does not reproduce its temporal pattern. Table~\ref{tab:composition_all_samples} compares the results across all five fitting sample sizes. The relative counterfactual range is defined as $C_{\mathrm{range}} = 100\Delta\alpha_{\mathrm{comp}}/ \Delta\alpha_{\mathrm{actual}}$. This is a range comparison only, not a measure of statistically explained variance. Across all sample sizes, the composition-only range represents $23.65\%$--$34.53\%$ of the fitted overall range for USDT and $9.86\%$--$11.17\%$ for USDC. The consistency of these values across a fitting sample size confirms that the limited variation of the composition-only exponent is not an artifact of the choice of $N$.

In figure~\ref{fig:weight_alpha_space}, we provide an additional view by plotting the transaction weight $w_i(t)$ of each category against its exponent $\alpha_i(t)$. The EOA-involved categories are generally concentrated at lower exponent values at $\alpha \approx 1.45$--$1.60$, whereas SC-SC transactions tend to exhibit higher and more variable exponents with $\alpha \approx 1.72$--$1.73$. These differences remain visible despite changes in category weights across periods, indicating that the exponent is associated more strongly with the interaction mechanism than with transaction prevalence alone.  Overall, changes in transaction weights alone are insufficient to reproduce the observed variation of the overall exponent. The counterfactual analysis shows that composition changes account for only $10\%$-$35\%$ of the temporal range, and the weight-exponent space confirms that SC-SC and the EOA-involved categories occupy fundamentally distinct regions regardless of their weights. Together, these results point to persistent interaction-specific differences in tail behaviour, rather than shifts in transaction composition, as the primary driver of how the overall exponent evolves over time. However, a pooled-distribution exponent is not generally equal to a weighted average of its component exponents. The $\alpha_{\mathrm{comp}}$ should be interpreted as a descriptive counterfactual indicator rather than an exact decomposition of the overall exponent.

\begin{table}[h]
\centering
\caption{Table contains the temporal ranges of the directly fitted overall and composition-only counterfactual exponents across the six study periods. The relative range, $100\Delta\alpha_{\mathrm{comp}}/\Delta\alpha_{\mathrm{actual}}$, remains consistently small ($10\%$--$35\%$) across all fitting sample sizes, demonstrating that composition changes alone do not reproduce the observed temporal variation of the overall exponent.}
\label{tab:composition_all_samples}

\begin{tabular}{lcccc}
\toprule
Token & Sample size
& $\Delta\alpha_{\mathrm{actual}}$
& $\Delta\alpha_{\mathrm{comp}}$
& Relative range (\%) \\
\midrule
USDT & $50{,}000$  & $0.0767$ & $0.0181$ & $23.65$ \\
     & $100{,}000$ & $0.0600$ & $0.0177$ & $29.48$ \\
     & $200{,}000$ & $0.0581$ & $0.0185$ & $31.80$ \\
     & $400{,}000$ & $0.0577$ & $0.0175$ & $30.39$ \\
     & $800{,}000$ & $0.0506$ & $0.0175$ & $34.53$ \\
\midrule
USDC & $50{,}000$  & $0.2199$ & $0.0246$ & $11.17$ \\
     & $100{,}000$ & $0.2544$ & $0.0256$ & $10.08$ \\
     & $200{,}000$ & $0.2583$ & $0.0265$ & $10.27$ \\
     & $400{,}000$ & $0.2585$ & $0.0255$ & $9.86$ \\
     & $800{,}000$ & $0.2488$ & $0.0264$ & $10.61$ \\
\bottomrule
\end{tabular}
\end{table}

\par\medskip

\noindent
{\centering

\includegraphics[width=0.96\textwidth]{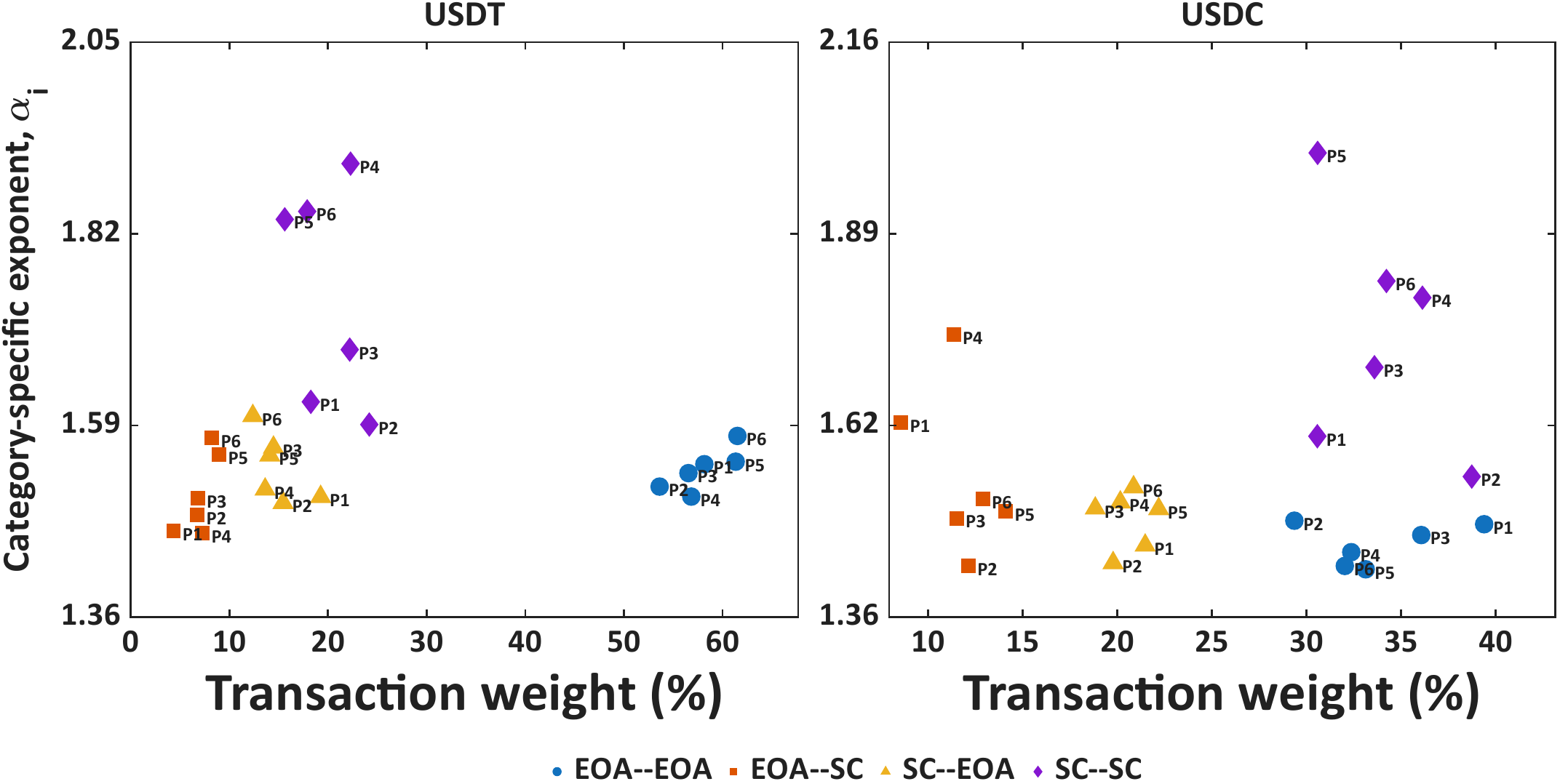}

\par}

\refstepcounter{figure}
\label{fig:weight_alpha_space}

\par\smallskip

\noindent
\textbf{Fig. \ref{fig:weight_alpha_space}.}
Relationship between transaction weight and the category-specific exponent, $\alpha_i$, for USDT and USDC at $N=200{,}000$. Each marker represents one study period, with exponents averaged across five random seeds. P1--P6 denote the six periods.

\par\medskip

\section{Conclusion}
\label{sec:Conclusion}

This study investigated whether stablecoin transaction values exhibit universal scaling behaviour or whether their statistical properties depend on the underlying blockchain interaction mechanism. We analysed approximately 370 million USDT and USDC transactions recorded on the Ethereum blockchain across six study periods. Based on the interaction types between Externally Owned Accounts (EOAs) and Smart Contracts (SCs), we classified the transactions into four categories, namely EOA-EOA, EOA-SC, SC-EOA, and SC-SC. Our results show that the transaction value distributions exhibit heavy-tailed power-law scaling across both stablecoins, all six study periods, and all interaction categories. The estimated exponents fall mainly within the range $\alpha\approx1.4$--$1.8$, broadly consistent with the heavy-tailed scaling previously reported for financial-market quantities such as trade sizes and share volumes \cite{stanley2008statistical}. Furthermore, we identify two interaction-specific scaling regimes. The EOA-involved categories cluster at average exponents, $\alpha\approx1.45$--$1.60$, whereas SC-SC transactions exhibit higher average exponents of approximately $\alpha\approx1.72$--$1.73$. This separation remains broadly robust across the two stablecoins, study periods, and fitting sample sizes. These results indicate that the scaling exponent is systematically associated with the underlying interaction mechanism. The counterfactual analysis further shows that changes in the relative transaction weights of the four categories alone are insufficient to reproduce the magnitude or temporal pattern of the overall exponent.

Together, these findings identify distinct and robust scaling regimes in stablecoin blockchain transaction value in the Ethereum blockchain. The sensitivity analysis shows that the estimated exponents and their temporal patterns remain stable across fitting sample sizes $N$ = $50{,}000, 100{,}000, 200{,}000, 400{,}000, 800{,}000$ and across five random seeds. The composition analysis shows that changes in the relative transaction shares of the four categories do not reproduce the magnitude or temporal pattern of the directly fitted Overall exponent. Thus, the overall exponent represents a pooled outcome of heterogeneous interaction categories and can hide important differences between EOA-involved and fully smart contract transfers. These findings show that the power-law behaviour is observed across all transaction types, but the exponent itself changes depending on who is transacting—whether it's EOAs or Scs.

The present analysis is limited to USDT and USDC transactions on the Ethereum blockchain. Taken together, these results do not establish a single universal tail exponent for stablecoin transactions. They show broadly recurring power-law-like tails together with robust interaction-specific differences in the fitted exponents. These patterns may indicate conditional scaling regularity within transaction classes, but confirmation of universality would require formal tail-model validation and evidence from additional stablecoins, blockchains, and market conditions. Future work will extend our work to other stablecoins, such as DAI, BUSD, and USDP, and to other blockchain platforms, including Binance Smart Chain, Solana, and Polygon. Such an extension allows us to verify whether the power-law tail behavior observed here holds across different blockchain ecosystems, making the findings more robust and generalizable.

\section*{Declaration of competing interest}
The authors declare that they have no known competing financial interests or personal relationships that could have appeared to influence the work reported in this paper.

\section*{Acknowledgment}

The authors gratefully acknowledge the National Institute of Technology Sikkim for providing a doctoral research fellowship to Kundan Mukhia. The authors also acknowledge the Brahmagupta High-Performance Computing Centre, Sikkim University, for providing the computational facilities used for data analysis. We gratefully acknowledge Dr. Amitabha Bhattacharyya, Head, Department of Physics, and Dr. Rupak Mukherjee, Department of Physics, Sikkim University, for facilitating access to these facilities. The authors also thank Britan Singh for his valuable assistance and support during this work.

\section*{Data availability}
Data is publicly available. The link is given in the paper~\cite{xblockwebsite}. The preprocessed data used for the analysis are available from the author upon request.

\printcredits

\bibliographystyle{model1-num-names}

\bibliography{cas-refs}

\end{document}